A.M.Galper, B.I.Luchkov, Yu.T.Yurkin

*National Research Nuclear University MEPhI*
*Moscow, Kashirskaya Road 31*

# VARIABLE SOURCE of HIGH ENERGY GAMMA-RAY RADIATION CYGNUS X-3

**Abstract.** Long observation (1972 – 2009) of the powerful discrete source in Galaxy Cygnus X-3 discovered its main properties and let to develop a real model of this unique object. It is short binary system with 4.8 hour orbital period including relativistic object (neutron star or black hole) and massive star. Source most activity is seen in gamma-rays from tens MeV to thousands TeV.

In 1970s during three measurements aboard high-altitude balloons we discovered sharp variable gamma-ray radiation with energy E ≥ 40 MeV from famous source Cygnus X-3 [1, 2]. The device was multi-gap spark chamber surrounded by anti-coincident counter for charged particles rejection. In chamber lead plates gamma-rays converted to $e^+$-$e^-$ pairs, multiple scattering served to measure gamma-ray energy. Most intensive gamma-ray flux was observed in 1972 after strong radio-flare of the source. On the same time a team of Crimea Astronomical Observatory measured its gamma-ray flux with energy E > 1 TeV by means of overland gamma-telescope registered cherenkov light of atmospheric shower [3, 4]. Joint analysis gave rise to find intrinsic period P = 4.8 hour, which is binary system orbital period, burst phases being equal in both spectral intervals (≥ 40 MeV and ≥ 1 TeV). The source also was observed at greater energies ( ~ $10^3$ TeV) by land arrays for large atmospheric showers [6, 7]. Integral energy spectrum is shown in fig. 1 [5] and can be approximated by single dependence with index α = 1.2 ± 0.3 up to energy ~ $10^{16}$ eV.

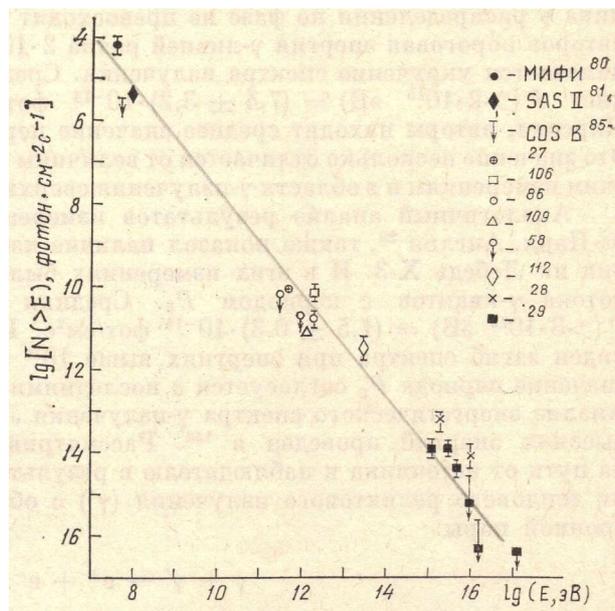

As conclusion Cygnus X-3 appears unique highly energetic object in the Galaxy, natural particle accelerator which could produce most part of cosmic rays in wide energy range $10^8 - 10^{16}$ eV [5].

Now times (2008-2009) orbital devices Agile [8] and Fermi [9] registered Cygnus X-3 gamma-ray emission of energies ~ 100 MeV with strong 4.8 hour modulation. The source is assumed to be binary system, consisting of microquasar (accretion neutron star or black hole) and Wolf-Rayet star (tens solar mass object on end evolutionary state). In these observations, as 40 years ago, great gamma-ray activity was correlated with strong radio flares in wide length wave band.

So far there are no new observations in ultrahigh energies ( > 1 TeV and > 10 TeV).

Comparison of our data with new obtained results confirms that Cygnus X-3 indeed presents high energy flaring object gamma-ray power of which changes for many orders depending of compact component activity phase, which nature is not yet entirely found [10]. Microquasar Cygnus X-3 exploration, being not so far (~ 12 Kpc), let us understand better farway quasars at cosmological distances (> $10^3$ Mpc). Therefore Cygnus X-3 investigation is very important because of valuable information for cosmology.